\documentclass[12pt]{iopart}

\usepackage{graphicx}

\begin{document}
\title[Cosmological models of coupled scalar
and Abelian gauge fields]{Applications 
to cosmological models of a complex scalar field
coupled to a $U(1)$ vector gauge field}

\author{Daniele S. M. Alves\dag\ and Gilberto M. Kremer\dag  
\footnote[3]{To
whom correspondence should be addressed (kremer@fisica.ufpr.br)}
}

\address{\dag\ Departamento de F\'\i sica, Universidade Federal do Paran\'a,
Caixa Postal 19044, 81531-990 Curitiba, Brazil}

\def\be{\begin{equation}}
\def\ee#1{\label{#1}\end{equation}}
\def\sqr#1#2{{\vcenter{\vbox{\hrule height.#2pt
\hbox{\vrule width.#2pt height#1pt\kern#1pt
\vrule width.#2pt}
\hrule height.#2pt}}}}
\def\qq{{\mathchoice\sqr{17.2}4\sqr{16.4}4\sqr53\sqr{11.6}3}}

\def\qs{{\mathchoice\sqr{7.2}4\sqr{6.4}4\sqr53\sqr{2.6}3}}

\begin{abstract}
We consider the Abelian model of a complex scalar field coupled to a gauge 
field within the framework of General Relativity and search for cosmological 
solutions. For this purpose we assume a homogeneous, isotropic and uncharged 
Universe and a homogeneous scalar field. This model may be inserted in several 
contexts in which the scalar field might act as inflaton or quintessence, 
whereas the gauge field might play the role of radiation or dark matter, 
for instance. Particularly, we propose two such models: (i) in the first, 
the inflaton field decays to massive vector bosons that we regard as 
dark-matter; (ii) in the second, due to its coupling to radiation the scalar 
field is displaced from its ground state and drives an accelerated expansion 
of the Universe, playing the role of quintessence. We observe that the 
equations are quite simplified and easier to be solved if we assume a 
roughly monochromatic radiation spectrum. 
\end{abstract}



\maketitle

\section{Introduction}

The study of scalar fields in cosmological models has been proven to be very 
fruitful both in inflationary theories and in quintessence 
models.~\cite{inflation,Kremer,Linde,KT,revq,quint}

It turns out that the success of these models also raises the need to 
investigate the interaction of these scalar fields with other forms of matter.

The inflationary theories are based upon the dynamical evolution of a 
scalar field
whose coupling to other fields is normally ignored until reheating 
process takes place~\cite{inflation,Linde,KT,reheating}. 
In the first reheating models a phenomenological friction 
term $\Gamma\dot\phi$ was inserted into the equation of motion of the scalar 
field in order to simulate the energy transfer to other forms of 
matter~\cite{gammaphiponto}. The theories of reheating have been 
improved since then, and today they are very 
more sophisticated, taking under consideration quantum effects and many kinds 
of possible decays of the inflaton to other particles~\cite{ressonance,KLS}. 
Anyway, how reheating 
really occurred is still an object of intense investigation~\cite{Kof}.

The very recent measurements of type-IA supernovae~\cite{SNIa} 
and the analysis of the 
power spectrum of the CMBR~\cite{CMBR} provided strong evidence for a 
present accelerated expansion of the Universe~\cite{Carroll,accel}. 
That offered a new setting for the scalar field 
cosmological models. In these contexts they are named quintessence 
fields~\cite{revq}, 
and recently many interesting phenomenological models of interacting 
quintessence have been proposed~\cite{interacq}.

In the present work a very simple cosmological model is investigated in which 
a complex scalar field interacts with a vector field. We believe that the kind 
of interaction treated in this model is very special because it raises 
naturally by requiring a local $U(1)$ gauge symmetry. 
This is the principle 
beyond the electromagnetic interaction and it inspired the development of 
the other fundamental interactions. 
According to the current view all the 
forms of interactions are completely determined by certain internal symmetry 
groups~\cite{qft}.

Our starting point will be the following Lagrangian:
\be
{\cal L}_{\rm mat}=D_\mu\phi(D^\mu\phi)^*-V(\phi\phi^*)
-{1\over 4}F_{\mu\nu}F^{\mu\nu},
\ee{1}
where 
\be
D_\mu\equiv\nabla_\mu+iqA_\mu,\qquad F_{\mu\nu}\equiv\partial_\mu A_\nu-
\partial_\nu A_\mu.
\ee{1a}
The Lagrangian (\ref{1})
is clearly invariant under local $U(1)$ gauge transformations:
$\phi\rightarrow \exp[{iq\alpha(x)}]\phi$ and
$A_\mu\rightarrow A_\mu -\partial_\mu\alpha(x)$, and 
associated with this symmetry is the conservation of the four-current
\be
{\cal J}^\mu=iq[\phi (D^\mu\phi)^*-\phi^* (D^\mu\phi)].
\ee{1b}

Our basic procedure will be to introduce the Lagrangian (\ref{1}) in the 
action with the Einstein-Hilbert Lagrangian and derive the Einstein's 
equations 
for that system (Sec.2). Then we shall investigate solutions to those 
equations that may be applicable in cosmological models. In Sec.3 we 
ascribe an interpretation to the contribution of each field for the energy 
density. We also discuss about the flow of energy among the fields. The 
Fourier expansion of the vector field and spatial average procedures are 
performed in Sec.4 as a first step towards the application to concrete 
models. Considerations about the possibilities of such applications comes 
in Sec.5, and the equations are solved for the simplified case of a roughly 
monochromatic radiation spectrum. Finally, 
we conclude in Sec.6 with brief 
remarks and future prospects. All the fields are treated 
classically. We choose units such that $c=\hbar=k=1$ and adopt the convention 
that Greek indices run from $0$ to $3$ whereas Latin indices run from $1$ to 
$3$.

\section{Field equations}

As was previously mentioned, this model can describe the interaction of the
scalar field with the electromagnetic field, provided we identify the gauge 
vector field with the electromagnetic four-potential~\cite{qft}. 
For the time being, that 
is what we shall bear in mind in our procedure. However, in later sections 
we will investigate cases in which spontaneous symmetry breaking occurs, so 
it is convenient to keep an open mind in the sense that the vector field 
$A_\mu$ does not necessarily need to be identified with the electromagnetic 
vector potential.

Let us consider the action:
\be
S=\int d^4x\sqrt{-g}\left[{R\over 16\pi G}+D_\mu\phi(D^\mu\phi)^*
-V(\phi\phi^*)-
{1\over4}F_{\mu\nu}F^{\mu\nu}\right].
\ee{2}
By varying the action with respect to $\phi^*$, $\phi$ and $A_\mu$, 
we obtain the equations:
\be
D_\mu D^\mu\phi=-{\partial V\over \partial \phi^*},\qquad
(D_\mu D^\mu\phi)^*=-{\partial V\over \partial \phi},
\qquad
{F^{\mu\nu}}_{;\nu}=-{\cal J}^\mu,
\ee{3}
respectively. By the anti-symmetry of $F^{\mu\nu}$ and the 
symmetry of $R_{\mu\nu}$, eq.(\ref{3})$_3$ implies that 
${{\cal J}^\mu}_{;\mu}=-{F^{\mu\nu}}_{;\nu;\mu}=-R_{\mu\nu}F^{\mu\nu}=0$.

The Einstein's field equations follow by requiring a stationary action with 
respect 
to variations of the metric $g_{\mu\nu}$, and turn out to be:
\be
G_{\mu\nu}=-8\pi G\left\{T_{\mu\nu}^{\rm rad}+2D_{(\mu}\phi[D_{\nu)}
\phi]^*
-g_{\mu\nu}\left[D_\sigma\phi(D^\sigma\phi)^*-V(\phi\phi^*)\right]
\right\},
\ee{4}
where $T_{\mu\nu}^{\rm rad}$ is the well-known Maxwell energy-momentum tensor 
of the electromagnetic field~\cite{elmag}: 
$T_{\mu\nu}^{\rm rad}={F_\mu}^\sigma 
F_{\sigma\nu}+g_{\mu\nu}F_{\alpha\beta}F^{\alpha\beta}/4$.

We are interested in solutions to those equations that describe a homogeneous 
scalar field interacting with a homogeneous and isotropic fluid of incoherent 
radiation. In that purpose, we will stress some points and make some 
considerations:

(i) the space-time geometry will be that of a plane Robertson-Walker metric, 
$ds^2=dt^2-a(t)^2(dx^2+dy^2+dz^2)$;

(ii) we will assume a homogeneous scalar field, $\phi=\phi(t)$;

(iii) we will describe an uncharged Universe, such that the charge density 
${\cal J}^0=0$;

(iv) the above conditions together with ${{\cal J}^\mu}_{;\mu}=0$ imply
that $\partial_iA^i=0$ . But that does not completely determine the field 
$A^\mu$, and we still have the gauge freedom to choose $A^0=0$ (Coulomb gauge);

(v) the choice $A^0=0$ and the $\mu=0$ component of eq. (\ref{3})$_3$ 
imply that 
$\dot\phi\phi^*-\phi{\dot\phi}^*=0$. By writing $\phi$ as $\phi_1+i\phi_2$, 
we see that the last equality means that $\phi_2=\alpha\phi_1$, and that 
allows us to define the real field $\Phi$ such that $\Phi^2\equiv\phi\phi^*
=\phi_1^2+\phi_2^2=(1+\alpha^2)\phi_1^2$;

(vi) by considering the free-field Lagrangian ${\cal L}_\phi^{\rm free}=
\partial_\mu\phi\partial^\mu\phi^*-V(\phi\phi^*)$, we can use Noether's 
theorem to derive the canonical energy-momentum tensor for the scalar field, 
$S_{\mu\nu}^\phi=2\partial_{(\mu}\phi\partial_{\nu)}\phi^*-g_{\mu\nu}
(\partial_\sigma\phi\partial^\sigma\phi^*-V)$. Regarding $S_{\mu\nu}^\phi$ 
as the stress-tensor of a perfect-fluid, we can make the following 
identifications:
\be
\rho_\Phi={S_0}^0=\dot\Phi^2+V(\Phi^2),\qquad
p_\Phi=-{1\over3}{S_i}^i=\dot\Phi^2-V(\Phi^2),
\ee{5}
where $\rho_\Phi$ and $p_\Phi$ are the energy density and pressure of the 
scalar field, respectively;

(vii) we will assume that $T_{\mu\nu}^{\rm rad}$ is the energy-momentum tensor 
of a homogeneous and isotropic fluid of incoherent radiation. That means that 
in a co-moving system there is no average net flow 
of electromagnetic 
energy in any direction~\cite{Tolman,TE}, 
so that $T_{\mu\nu}^{\rm rad}$ is diagonal 
and $p_{\rm r}=\rho_{\rm r}/3$, where $p_{\rm r}$ is the radiation pressure 
and $\rho_{\rm r}$ is the radiation energy density given by $\rho_{\rm r}=
T_{00}^{\rm rad}$.

In view of the above considerations we can write the system of equations 
for $a(t)$, $\Phi(t)$, and $A_i(x^\mu)$ (bearing in mind that $A_0=0$) as:
\be
H^2={8\pi G\over3}\left[\rho_{\rm r}+\rho_\Phi-q^2A_iA^i\Phi^2\right],
\ee{6}
\be
{\ddot a\over a}=-{4\pi G\over3}\left[\rho_{\rm r}+\rho_\Phi+3(p_{\rm r}
+p_\Phi)\right],
\ee{7}
\be
\ddot\Phi+3H\dot\Phi+{1\over2}{dV\over d\Phi}=q^2A_iA^i\Phi,
\ee{8}
\be
\partial^\mu\partial_\mu A_i+H\dot A_i=-2q^2\Phi^2A_i.
\ee{9}
Equations (\ref{6}) and 
(\ref{7}) refer to the two linearly independent Einstein's 
field equations (\ref{4}), whereas  (\ref{8}) and (\ref{9}) follow from 
(\ref{3}).

\section{Balance equations and interpretation}

It is well-known from elementary electromagnetic theory~\cite{elmag} 
that the magnetostatic 
energy of a current distribution is:  $U={1\over2}\int_V\mathbf{J}\cdot
\mathbf{A}dV$, where $\mathbf{J}$ is the current density and $\mathbf{A}$ is 
the magnetic vector potential. Hence, it is possible to associate an energy 
density ${1\over2}\mathbf{J}\cdot\mathbf{A}$ to a system of currents. 
Regarding to our system, in which ${\cal J}^i=-2q^2A^i\Phi^2$, the current 
energy density is given by 
\be
\rho_{\cal J}={1\over2}A_i{\cal J}^i=-q^2A_iA^i\Phi^2.
\ee{10}

Therefore, we can rewrite the Friedmann equation (\ref{6}) in its familiar 
form $H^2=8\pi G\rho/3$, and ascribe to it a simple physical 
interpretation: the total energy density of the system, $\rho$, is the 
sum of three forms:

(i) $\rho_{\rm r}$, the energy density of the electromagnetic field, 
which may also be regarded as the radiation energy;

(ii) $\rho_\Phi$, the energy density of the scalar field;

(iii) $\rho_{\cal J}$, the current energy density. Moreover, the current 
pressure is given by $p_{\cal J}=-\rho_{\cal J}/3$, as can be seen 
by identifying the energy-momentum tensor ${T_\mu}^\nu$ with that of a 
perfect fluid: ${T_\mu}^\nu={\rm diag}(\rho,-p,-p,-p)$, where $\rho=
\rho_{\rm r}+\rho_\Phi+\rho_{\cal J}$ and $p=p_{\rm r}+p_\Phi
+p_{\cal J}$.\footnote{It is interesting to note that, due to such
barotropic relation ($p_{\cal J}=-\rho_{\cal J}/3$),
the current energy density does not contribute to the
acceleration, eq.(\ref{7}), since such a
contribution would show up in the form $\rho_{\cal J}
+3p_{\cal J}$, which is identically zero.}

Let us now write the balance equations for $\rho_{\rm r}$, $\rho_\Phi$, and 
$\rho_{\cal J}$. First from ${T^{\mu\nu}_{\rm rad}}_{;\nu}=-F^{\mu\nu}
{\cal J}_\nu$ it follows the balance equation for the energy density of the 
radiation field, i.e.,
\be 
\dot\rho_{\rm r}+3H(\rho_{\rm r}+p_{\rm r})=-{\cal J}^i{\dot 
A_i}.
\ee{11a}
We note that the term $-{\cal J}^i\dot A_i=-\mathbf{J}
\cdot\mathbf{E}$ represents the power flux transfered to the electromagnetic 
field due to the currents motion~\cite{elmag}.
Next, differentiation of $\rho_\Phi=\dot\Phi^2+V(\Phi^2)$ with respect 
to time together with (\ref{8}) lead to 
\be
\dot\rho_\Phi+3H(\rho_\Phi+p_\Phi)=2q^2A_iA^i\Phi\dot\Phi,
\ee{11b}
which is the balance equation for energy density of the scalar field. 
Finally, from the expression for $\rho_{\cal J}$, eq. (\ref{10}), it 
follows the balance equation for the energy density of the currents
\be
\dot\rho_{\cal J}+3H(\rho_{\cal J}+p_{\cal J})={\cal J}^i\dot A_i-2q^2A_iA^i
\Phi\dot\Phi.
\ee{11}
Summation of eqs. (\ref{11a}), (\ref{11b}) and (\ref{11}) leads to the 
balance equation for the total energy density of the system
\be
\dot \rho+3H(\rho+p)=0,
\ee{11c}
naturally not independent of the Friedmann equation (\ref{6}) and of the 
acceleration equation (\ref{7}).

The interactions among the fields might be depicted in the form of a diagram,
as shown below:
$$
\qq\!\!\!\! \!\!\!\!\rho_{\rm r}\;\;\buildrel{-{\cal J}^i \dot A_i}\over 
\longleftrightarrow\;\;\qq\!\!\!\! \!\!\!\!
\rho_{\cal J}\;\;
\buildrel{2q^2A^iA_i\Phi\dot\Phi}\over\longleftrightarrow\;\;
\qq\!\!\!\! \!\!\!\!\rho_{\Phi}
$$

\section{Fourier expansion and spatial average procedure}

We now turn to the problem of finding a solution to the system of equations 
(\ref{6}) through (\ref{9}). Since we will have to deal 
with a partial differential equation for the field $A_i$, our approach will 
be to expand it in a Fourier series and make some simplifying assumptions in 
order to make the analysis manageable. All of these assumptions turn out to be 
quite reasonable, and as we shall argue later, are quite well justified.

The expansion for the vector potential in a triple Fourier series can be 
written as:
\be
A_i(t,\mathbf{x})=\sum_{\mathbf{k}} M_i^\mathbf{k}(t) e^{i\mathbf{k}\cdot 
\mathbf{x}},
\ee{12}
where the coefficients in the expansion are related by $M_i^\mathbf{-k}=
({M_i^\mathbf{k}})^*$, since $A_i$ is real. From the condition 
$\partial_iA^i=0$ 
it follows that for each $\mathbf{k}$, $k_iM_i^{\mathbf{k}}=0$, where $k_i$ 
refers to the components of the wavevector $\mathbf{k}$.

The equation satisfied by the coefficients $M_i^{\mathbf{k}}(t)$ follows from 
(\ref{9}), and reads:
\be
\ddot M_i^{\mathbf{k}}+H\dot M_i^{\mathbf{k}}+{k^2\over a^2}M_i^{\mathbf{k}}
=-2q^2\Phi^2M_i^{\mathbf{k}}.
\ee{13}

Now, we argue that an average procedure is needed in order to describe a 
homogeneous and isotropic radiation field~\cite{Tolman,TE}. 
To that end we define the volumetric spatial average~\cite{Novello} of a 
quantity $\psi$ at time $t$ by:
\be
\overline{\psi}={1\over V}\int_V\psi\sqrt{-g}d^3x,
\ee{14}
where $V\equiv L^3$ is the volume of a fundamental cube.

Imposing periodic boundary conditions, the summation in expression (\ref{12}) 
extends over the discrete spectrum of modes of the wavevector $\mathbf{k}$ 
whose components run through the values $k_i={2\pi n_i/ L}$,  $n_i$ assuming 
integer values. In the limit that $L$ is large compared to the length scale 
in which homogeneity holds, the passage to the continuum description is well 
justified.

In the light of this we shall replace the expressions for $A_iA_i$ 
and $\rho_{\rm r}=T^{\rm rad}_{00}$ by its averaged values in 
order to make explicit their 
exclusive dependence on the variable $t$. In this case from
\be
A_iA_i=\left[\sum_{\mathbf{k}} M_i^\mathbf{k}(t) 
e^{i\mathbf{k}\cdot 
\mathbf{x}}\right]\left[\sum_{\mathbf{k'}} M_i^\mathbf{k'}(t) e^{i\mathbf{k'}
\cdot \mathbf{x}}\right],
\ee{15a}
\be
\rho_{\rm r}={1\over{2a^2}}\left[\dot A_i\dot A_i+{1\over a^2}
(\partial_iA_j\partial_iA_j-\partial_iA_j\partial_jA_i)\right],
\ee{15b}
we get the averaged values
\be
\overline{A_iA_i}(t)=\sum_{\mathbf{k}}M_i^{\mathbf{k}}(t)
{M_i^{\mathbf{k}}}(t)^*,
\ee{15}
\be
\overline{\rho_{\rm r}}(t)={1\over2a(t)^2}\sum_{\mathbf{k}}
\left[\dot M_i^{\mathbf{k}}(t){\dot M}_i^{\mathbf{k}}(t)^*
+{k^2\over a(t)^2}
M_i^{\mathbf{k}}(t){M_i^{\mathbf{k}}}(t)^*\right].
\ee{16}

We remark that the average procedure adopted here, besides making 
$A_iA_i$ 
and $\rho_{\rm r}$ functions of time $t$ only, accounts for the vanishing of 
the off-diagonal components of the energy-momentum tensor $T_{\mu\nu}$, which 
is required for self-consistency since the Einstein tensor $G_{\mu\nu}$ is 
diagonal in our homogeneous and isotropic geometry.

Now, since $M_i^{\mathbf{k}}(t)$ is a complex quantity, we will write it in 
its polar form:
\be
M_i^{\mathbf{k}}(t)=P_i^{\mathbf{k}}(t) e^{i\omega_i^{\mathbf{k}}(t)},
\ee{a}
and the equations for the real coefficients $P_i^{\mathbf{k}}(t)$ and 
$\omega_i^{\mathbf{k}}(t)$ can be obtained by inserting (\ref{a}) into 
(\ref{13}). Hence it follows
\be
\ddot\omega_i^{\mathbf{k}}+H\dot\omega_i^{\mathbf{k}}+2{\dot P_i^{\mathbf{k}}
\over P_i^{\mathbf{k}}}\dot\omega_i^{\mathbf{k}}=0,\qquad (i=1,2,3),
\ee{17}
which can be readily integrated and gives: 
\be
\dot\omega_i^{\mathbf{k}}={k\over a}{c_i^{\mathbf{k}}\over 
{P_i^{\mathbf{k}}}^2},
\ee{b}
where $c_i^{\mathbf{k}}$ is a constant. Moreover, 
\be
\ddot P_i^{\mathbf{k}}+H\dot P_i^{\mathbf{k}}+{k^2\over a^2}\left(P_i^{
\mathbf{k}}-{{c_i^{\mathbf{k}}}^2\over {P_i^{\mathbf{k}}}^3}\right)=-2q^2
\Phi^2P_i^{\mathbf{k}}.
\ee{18}

In terms of $P_i^{\mathbf{k}}(t)$ (and dropping bars), we have:
\be
A_iA_i(t)=\sum_{\mathbf{k}, i}{P_i^{\mathbf{k}}(t)}^2,
\ee{19}
\be
\rho_{\rm r}(t)={1\over 2a(t)^2}\sum_{\mathbf{k}, i}\left[{\dot{ P}_i^{
\mathbf{k}}(t)}^2
+{k^2\over a(t)^2}\left({P_i^{\mathbf{k}}(t)}^2+{{c_i^{
\mathbf{k}}}^2\over {P_i^{\mathbf{k}}(t)}^2}\right)\right].
\ee{20}

In short, we have the system of equations (\ref{7}), (\ref{8}), and (\ref{18}) 
to be solved for the functions $a(t)$, $\Phi(t)$ and the infinitely many 
$P_i^{\mathbf{k}}(t)$ coefficients. Solving for each $P_i^{\mathbf{k}}(t)$ 
means to find out how each frequency band of the radiation spectrum interacts 
and evolves. This task is certainly worth being investigated, and will be
performed in a future work.
Here, for the sake of simplicity we shall solve the
problem for the particular case in which all the
frequencies can be ignored except a specific one,
i.e., for the case of a monochromatic spectrum. This
procedure, to be considered in the next section, will
make the analysis more tractable and will allow us to
gain insight on the qualitative dynamical behavior of
the model.

\section{Monochromatic spectrum}

For a monochromatic spectrum, the system of equations (\ref{7}), (\ref{8}), 
and (\ref{9}) is simplified and reads:
\be
{\ddot a\over a}=-{4\pi G\over3}\left[\rho_{\rm r}+\rho_\Phi+3(p_{\rm r}
+p_\Phi)\right],
\ee{21}
\be
\ddot\Phi+3H\dot\Phi+{1\over2}{dV\over d\Phi}=-{q^2\over a^2} A_iA_i\Phi,
\ee{22}
\be
 \partial_0^2(A_iA_i)+ H\partial_0(A_iA_i)=
-4q^2\Phi^2A_iA_i+4a^2\rho_{\rm r}-4{k^2\over a^2}A_iA_i,
\ee{23}
where $\rho_\Phi=\dot\Phi^2+V$, $p_\Phi=\dot\Phi^2-V$, $\rho_{\cal J}=q^2
A_iA_i\Phi^2/a^2$ and $\rho_{\rm r}$ is given by the Friedmann equation, 
namely,
\[
H^2={8\pi G\over3}(\rho_{\rm r}+\rho_\Phi+\rho_{\cal J}),
\]
and is related to its pressure by $p_{\rm r}=\rho_{\rm r}/3$.

That system describes a homogeneous and 
uncharged Universe filled with monochromatic radiation coupled to a homogeneous
scalar field. Once we have chosen an appropriate potential 
$V(\Phi^2)$ and the initial amount of energy of  each field, that system 
will yield a unique 
solution describing the subsequent evolution of the Universe and its 
constituents.

Now it is an opportune moment to discuss the choices for the scalar field 
potential $V(\Phi^2)$. Up to now we have been identifying the vector field 
$A_\mu$ with the electromagnetic vector potential. The scalar field potentials 
$V(\Phi^2)$ that fit in this case are those which have a minimum at $\Phi=0$.
However, potentials whose lowest state is not at $\Phi=0$ might not be suitable
as concerns to radiation, since the symmetry will be spontaneously broken and 
the ''photons'' will become massive. Therefore, we will distinguish between 
two classes of potentials $V(\Phi^2)$: the ones with non-degenerate minimum 
and the rest.

In what follows we shall investigate the solutions to equations (\ref{21}), 
(\ref{22}) and (\ref{23}) in two contexts: $\emph{reheating}$ (subsection 5.1) 
and $\emph{quintessence}$ (subsection 5.2).

\subsection{Reheating and Dark-Matter}

Among the several situations in which this model may be applied, perhaps the 
most straightforward one is the period that succeeds the end of inflation, 
when the energy stored in the inflaton field decays to radiation and reheats 
the Universe. The most suitable potentials $V(\Phi^2)$ in this case are 
$V\propto\Phi^2$ or $V\propto\Phi^4$. Investigation of eqs. (\ref{21}) to 
(\ref{23}) in this context does not reveal any new feature than those already 
known from the literature in reheating processes~\cite{Kof,KLS}. 
In fact, this description 
faces the very difficulties that decaying oscillating scalar fields display, 
namely, that the decay products of the scalar field are ultrarelativistic and 
their energy densities decreases due to expansion much faster than the energy 
of the oscillating field, which goes as $\propto 1/a^3$. Therefore the energy 
stored in the inflaton field never decays completely, even when the coupling 
$q^2$ is large. As was remarked in~\cite{KLS}, reheating can be complete 
only if the 
decay rate of the scalar field decreases more slowly than $t^{-1}$. Typically 
this requires either spontaneous symmetry breaking or coupling of the inflaton 
to fermions with $m_\psi<m_\phi/2$. 

Nevertheless, this model also exhibits 
effects of resonance~\cite{KLS,ressonance} that may turn reheating efficient, 
although that requires a good
amount of fine tuning in the value of $k$ (see eq.(\ref{23})), that is, the 
system (\ref{21}), (\ref{22}) and (\ref{23}) would not be the most appropriate 
to allow for resonance behavior, and for a more accurate description
 the full spectrum 
of radiation must be taken into account.

\begin{figure}\begin{center}\vskip0.50cm
\includegraphics[width=7.5cm]{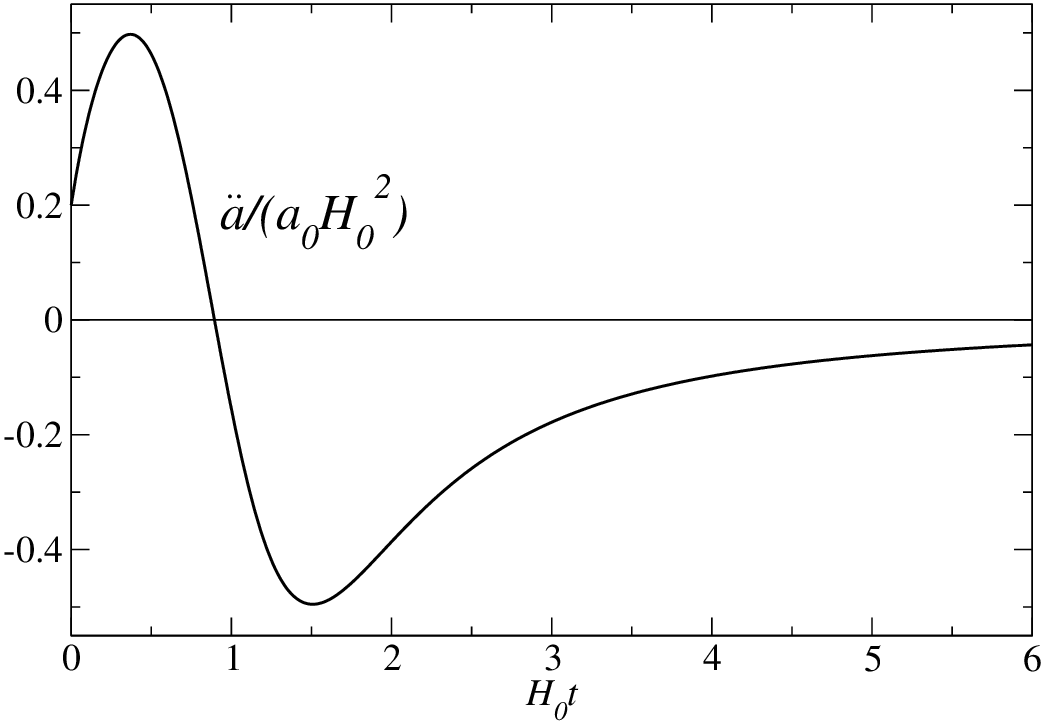}
\caption{The exit of inflation and the beginning of deceleration.}
\end{center}\end{figure}

Let us now lay radiation aside and allow the photons to gain mass. In that 
purpose, we will choose the potential $V(\Phi^2)=\mu
e^{-\lambda\Phi^2}$, where $\mu$ and $\lambda$ are constants. 
It is clear that, as $\Phi$ rolls down the potential hill 
towards non-zero values, the photons acquire an effective mass. Therefore this 
choice of potential can describe a scalar field that decays to massive vector 
bosons, which, in turn, can be responsible for dark matter in the Universe. To 
carry that out, we have to abandon our previous interpretation of the energy 
density of the fields which distinguish between $\rho_{\cal J}$ and 
$\rho_{\rm r}$, and ascribe to the massive vector field the energy density 
$\rho_{\rm dm}=\rho_{\rm r}+\rho_{\cal J}$ and pressure $p_{\rm dm}=p_{\rm r}
+p_{\cal J}$, where the subscript alludes to dark matter. In this redefinition,
$\rho_{\rm r}$ accounts for the kinetic part of the energy of the massive 
photons, while $\rho_{\cal J}$ owes to the contribution of its mass.

\begin{figure}\begin{center}\vskip0.50cm
\includegraphics[width=7.5cm]{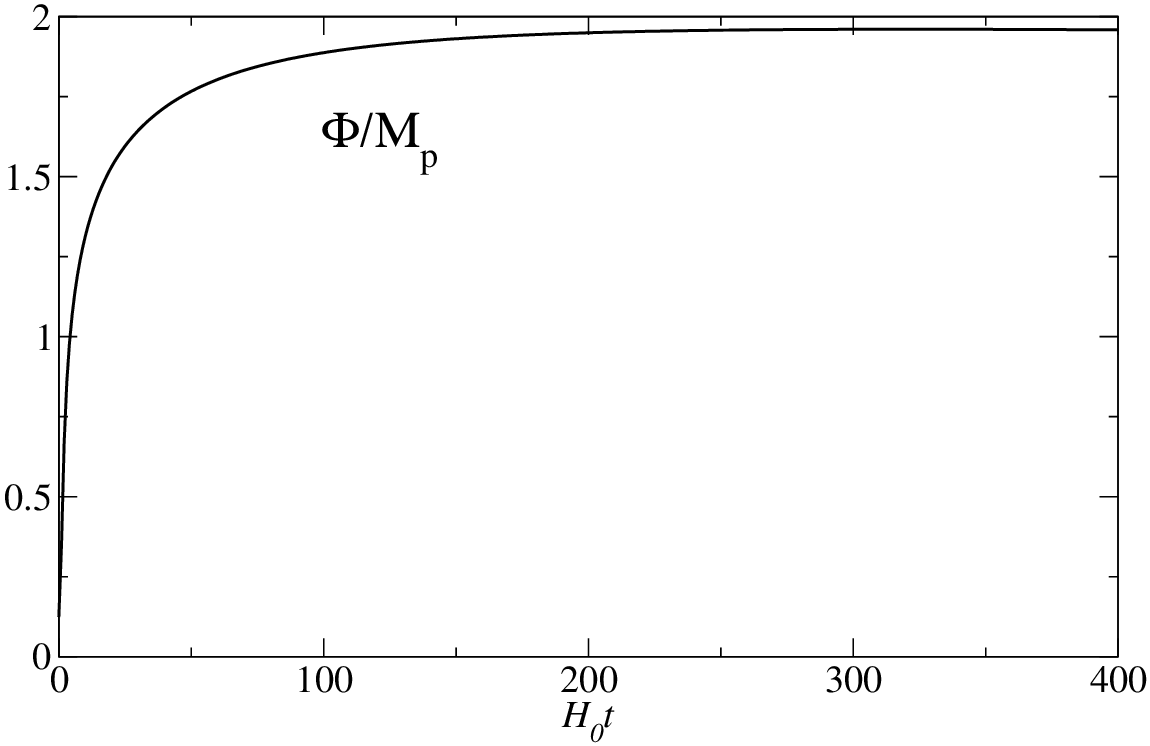}
\caption{The growth of the scalar field.}
\end{center}\end{figure}

By numerical analysis we extract the general features of the model. In figures 
1 to 3  we plot the relevant intervals for $\ddot a(t)$, $\Phi(t)$ and 
the energy 
densities of the fields for a particular choice of parameters:
\[k=2a_0H_0,\qquad q^2=0.05H_0^2M_p^{-2},\qquad\lambda=10M_p^{-2},\qquad 
\mu=0.7\rho_0,
\] 
and initial conditions:
\[ \cases{a(0)=a_0,\qquad \dot a(0)= a_0H_0,\qquad V(0)=0.6\rho_0,\cr
\dot\Phi(0)=0,\qquad
\rho_{\rm r}(0)=
k^2A_iA_i(0)/a_0^4,\qquad\partial_0(A_i A_i)(0)=0,}
\]
where $M_p$ is defined by $M_p\equiv\sqrt{3/(8\pi G)}$ and $\rho_0=H_0^2M_p^2$.
 However, the behavior
displayed is quite general and can be modeled by an adjustment of the 
parameters and initial conditions. It is described as follows: during 
inflation the scalar field is next to $\Phi=0$, where the potential 
$V(\Phi^2)$ is nearly flat and attains its maximal value. When the scalar 
field begins to depart from $\Phi=0$ and roll down the potential, inflation 
ends and the scalar field decays to massive vector bosons. The decay process 
is not abrupt; rather, there is a smooth transition from a domination of the 
scalar field to a domination of matter. Numerical solutions show that when 
the decay rate $2q^2A_iA^i\Phi\dot\Phi$ of the inflaton (see eq. (\ref{11b})) 
ceases to be significant, the scalar field attains a roughly constant value, 
$\Phi_{\rm c}$, implying an effective mass of the vector bosons $m_{\rm dm}=
\sqrt{2}q\Phi_{\rm c}$ (see eq. (\ref{9})), which, by this time, 
behave as ordinary matter with 
$p_{\rm dm}\approx0$ and $\rho_{\rm dm}$ scaling as $\propto 1/a^3$, whereas 
$\rho_\Phi$ decays faster than $\propto 1/a^4$. 

\begin{figure}\begin{center}\vskip0.50cm
\includegraphics[width=7.5cm]{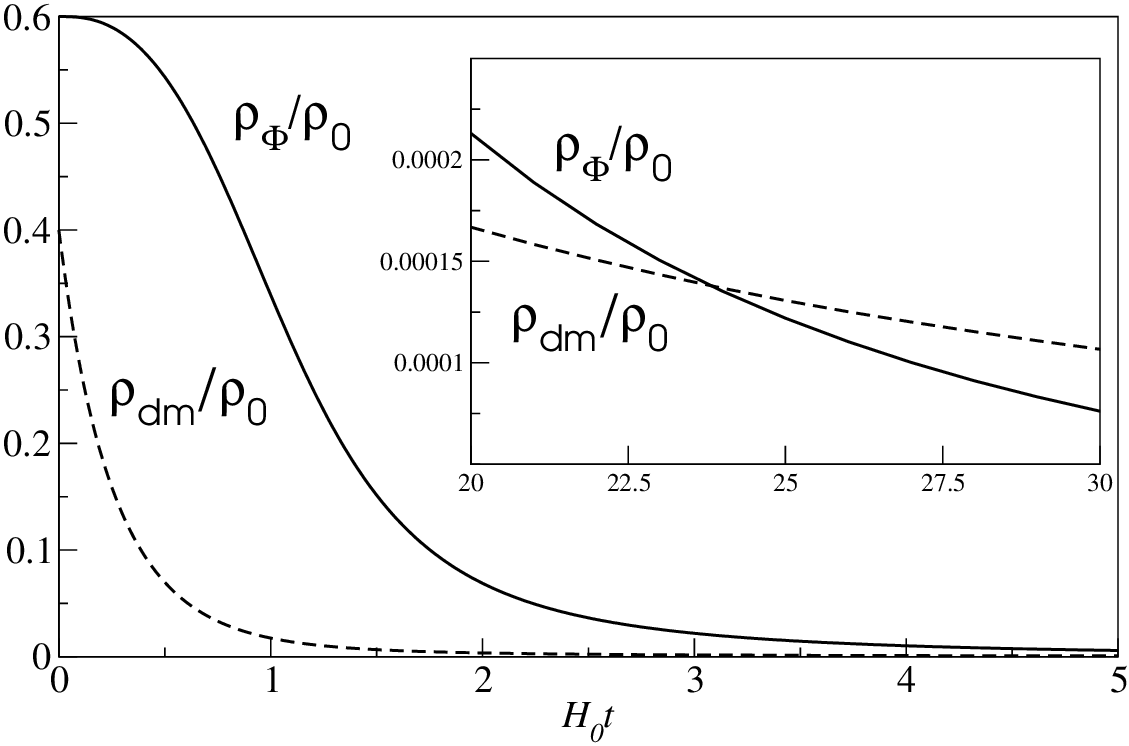}
\caption{The behavior of the energy densities of the dark matter 
(dashed line) and of the scalar field (straight line) and 
the smooth transition from the domination of the scalar field to the
domination of the massive vector field.}
\end{center}\end{figure}

\subsection{Quintessence}

By taking another look at the equation of motion for the scalar field:
\[
\ddot\Phi+3H\dot\Phi+{1\over2}V'=-{q^2\over a^2}A_iA_i\Phi
\]
we can spot a new term at the right hand side of the equality which is absent 
in the case of a non-interacting scalar field. This term shows up due to the 
interaction with the vector field, and it resembles an elastic force ($F=-kx$) 
whose time-dependent ``elastic constant'' is $q^2A_iA_i/a^2$. This force
acts on the motion of the scalar field by drawing it towards the position 
$\Phi=0$. When the scalar field is subject to a potential $V(\Phi^2)$ whose 
minimum lies at $\Phi=0$ the influence of this force will not alter 
significantly the qualitative behavior of the scalar field, but, on the 
contrary, if $\Phi=0$ is not a minimum, there must arise relevant 
modifications in its motion.

\begin{figure}\begin{center}\vskip0.50cm
\includegraphics[width=7.5cm]{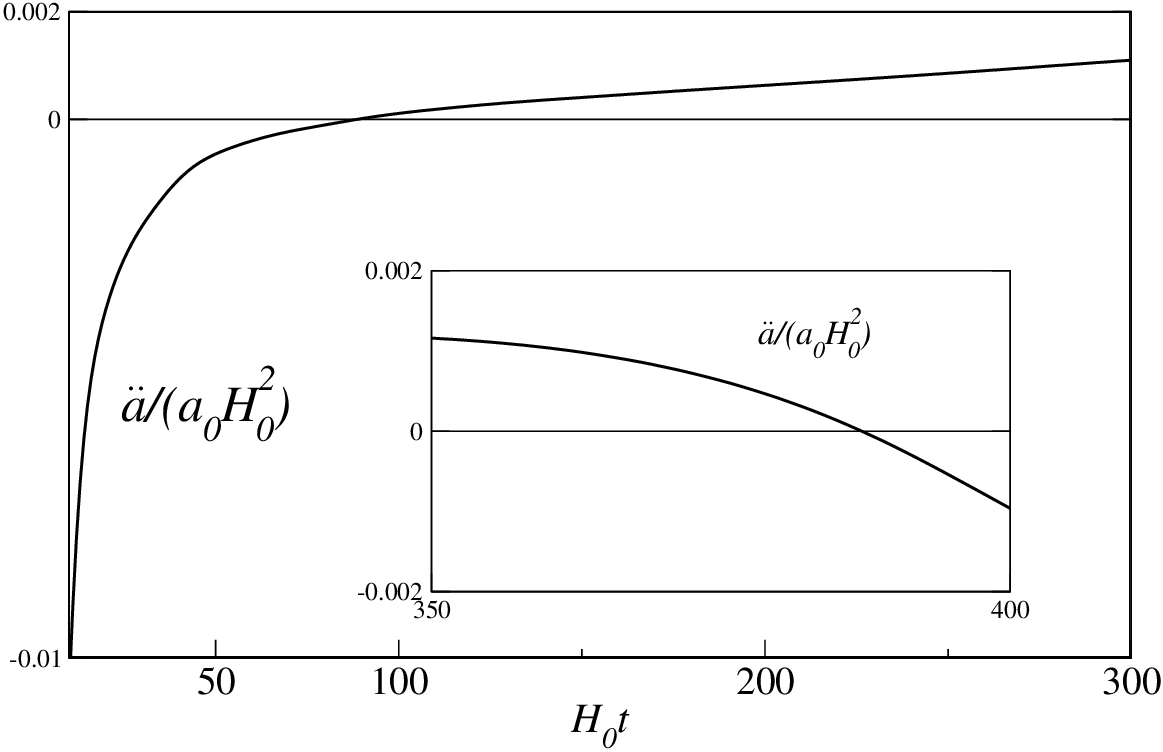}
\caption{Accelerated expansion driven by 
the quintessence field. In the small frame: end of quintessence era. }
\end{center}\end{figure}

That is what we shall investigate in this section by considering the 
``Mexican Hat'' potential 
\[
V(\Phi^2)=\lambda(\Phi^2-\Phi^2_0)^2, \qquad\hbox{where}
\qquad \Phi^2=\Phi^2_0
\]
corresponds to the  lowest state.  It is clear that the 
vacuum solution
$\Phi^2=\Phi^2_0$ and $A_\mu=0$
satisfies eqs. (\ref{8}) and (\ref{9}).

However, in the presence of vector bosons (i.e., if $A_iA_i\not
=0$), the 
ground state for the scalar field becomes an unstable position, for if we 
place $\Phi$ at $\Phi_0$, it will be pulled toward the origin due to the term 
$-q^2A_iA_i\Phi/a^2$.

\begin{figure}\begin{center}\vskip0.50cm
\includegraphics[width=7.5cm]{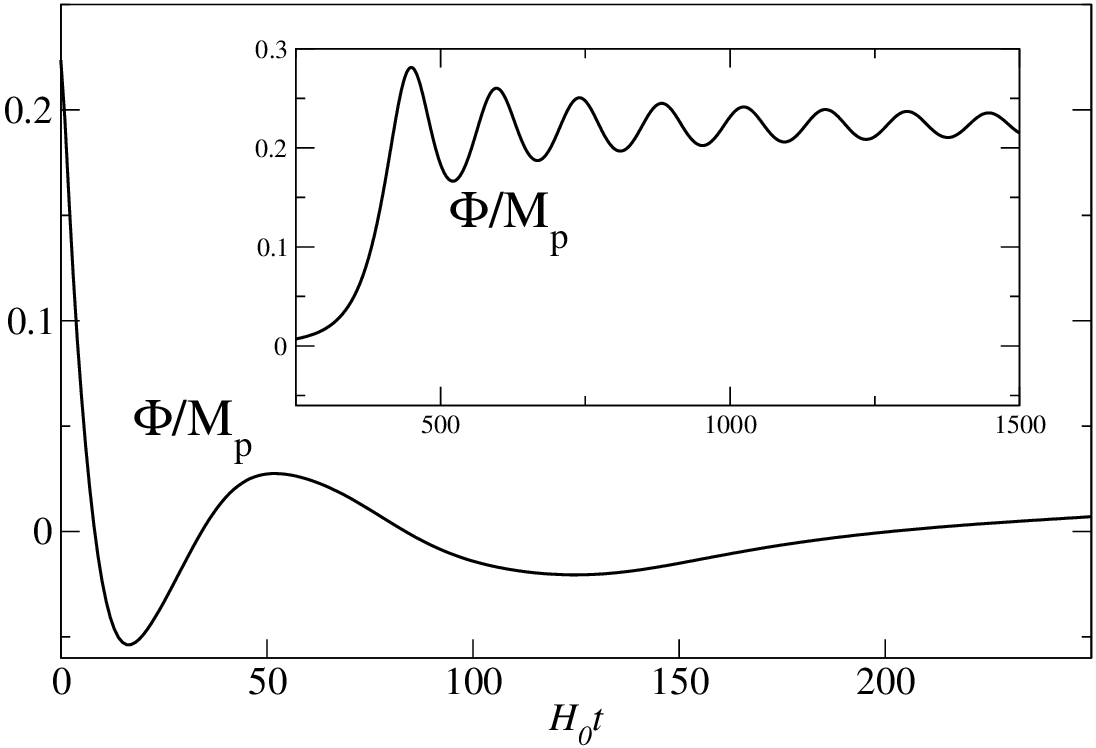}
\caption{The behavior of the field $\Phi$: (i) during the quintessence 
era; (ii)  in the subsequent period when it rolls back 
to its potential minimum and oscillates about it (small frame).}
\end{center}\end{figure}

With the following choice of parameters:
\[
k^2=0.7a_0^2H_0^2,\quad q^2=0.5H_0^2M_p^{-2}, 
\quad\lambda=0.01H_0^2M_p^{-2},\quad
\Phi_0^2=0.05M_p^2,
\]
we have solved numerically the system (\ref{21}), (\ref{22}) and (\ref{23}) 
with initial conditions
\[\cases{a(0)=a_0,\qquad \dot a(0)= a_0H_0,
\qquad\Phi(0)=\Phi_0, \cr
\dot\Phi(0)=0,\qquad 
\rho_{\rm r}(0)=
k^2A_iA_i(0)/a_0^4,\quad\partial_0(A_i A_i)(0)=0,}
\]
that corresponds to placing $\Phi$ in its ground state ($\rho_{\Phi}=0$) and 
filling the Universe with radiation. In figures 4 to 6 we plot the relevant 
intervals for $\ddot a(t)$, $\Phi(t)$ and the energy densities of the fields.

\begin{figure}\begin{center}\vskip0.50cm
\includegraphics[width=7.5cm]{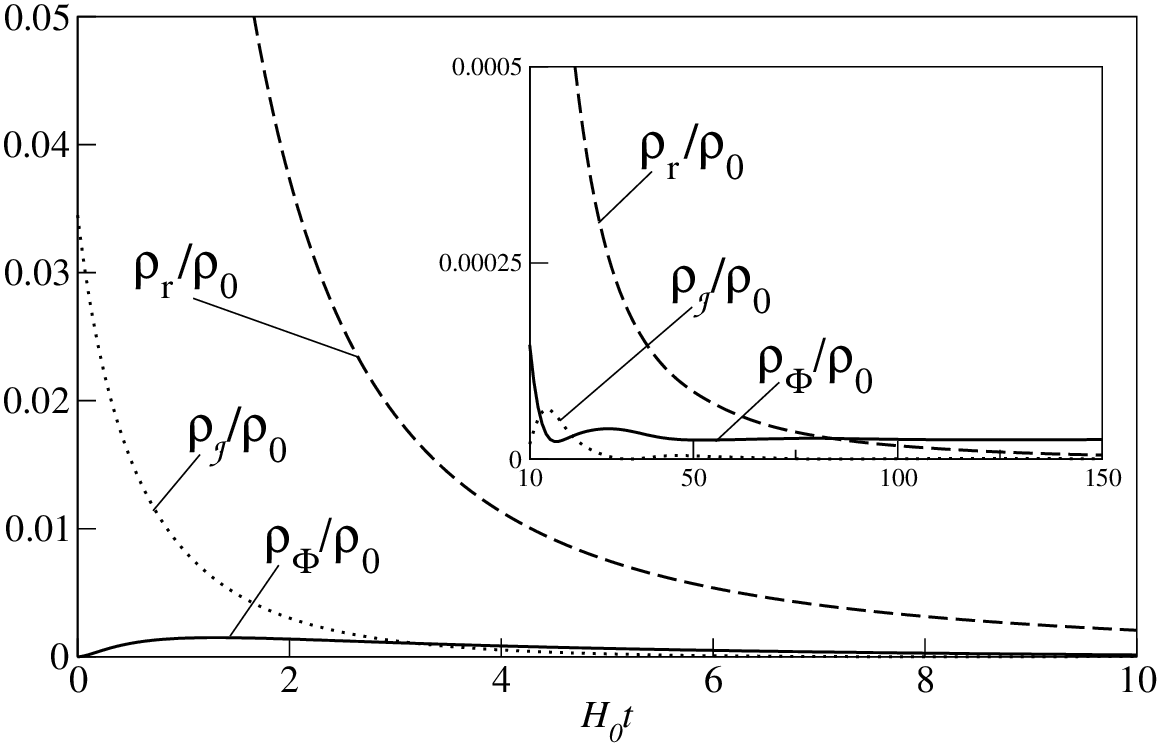}
\caption{The coupling to radiation transfers energy to the field $\Phi$, 
which later comes to dominate and behaves as quintessence (small frame).
Energy densities: scalar field -- straight line; current field --
dotted line; radiation field -- dashed line.}
\end{center}\end{figure}

The coupling of the scalar field to radiation removes it from its ground 
state and makes it climb up the potential hill to positions near the origin. 
That raises its potential energy and confers on it a quintessential behavior. 
During the period that $\Phi$ remains next to $\Phi=0$ the acceleration becomes
positive, $\rho_{\rm r}$ goes as $\propto 1/a^4$ and the vector bosons can be, 
meanwhile, regarded as radiation. This period can be made as long as one 
wishes by a suitable choice of the parameters. But, ultimately, the force that 
keeps $\Phi$ next to zero becomes too weak and $\Phi$ rolls back to (one of) 
its ground state(s), oscillating about it. Consequently, the symmetry is 
spontaneously broken, the photons become 
massive~\footnote{As was pointed out 
in the last subsection, when the symmetry is broken it is convenient to modify 
the interpretation of the energy densities, and ascribe to the massive photons 
the energy density $\rho_{\rm r}+q^2\Phi^2_0A_iA_i/a^2$, $\rho_{\rm r}$ 
being the kinetic contribution to its energy and $q^2\Phi^2_0A_iA_i/a^2$
the contribution due to its mass $m=\sqrt2q\Phi_0$.}, and the expansion 
decelerates again. 

As a final remark we note that, in this model, the positive acceleration of 
the Universe is not due to a small non-zero vacuum 
energy~\cite{Carroll,vacuum,revq}. Here, the vacuum 
energy is indeed zero, but the point is that the Universe is not in the vacuum 
state. It is filled with particles that, due to their interactions, displace 
the fields from their ground states and make them act as dark energy. Motivated
by this we suggest that a possible explanation for the present acceleration of 
the Universe might reside in the fact that the Universe is (obviously) not in 
vacuum although its energy density is very small, and the interactions among 
its constituents, which so far have been commonly neglected, might be 
responsible for the dark energy.

\section{Conclusions and Outlook}

To sum up, we remark that the qualitative treatment
performed here point out to interesting mechanisms of
accounting for the dark-matter in the Universe (or at
least a part of it), and also provides a possible
candidate for quintessence. However, the quintessence
behavior displayed here differs from the usual
quintessence models in the sense that the scalar field
is not tending asymptotically to the minimum of its
potential. Rather, its minimum is not a stable
position and the scalar field is removed from its
ground state due to its coupling to a background field
of vector bosons.\footnote{A similar mechanism was employed to 
produce a short secondary stage of inflation. See~\cite{felder} and
references therein.}

All the same, our study is not complete and a
quantitative confrontation of the predictions yield by
these models to observations is needful. Besides taking into account the full
spectrum, it is also necessary to address the issue of whether the models
treated here satisfy the constraints from the
anisotropy of the CMBR, the large-scale galaxy
distribution and the SN-Ia data. That will be
performed in a future work.

\ack
This work was supported by CNPq (Brazil).

\section*{References}



\end{document}